\documentclass{article}
\usepackage{hiph-preprint}
\volnumber{22} \issuenumber{1} \edyear{2005}                             
\frompage{000} \topage{000}                                              
\recrevdate{1 January 2005}                                              
\def\rightmark{$\pi^0$ measurement in $\sqrt{s_\mathrm{NN}}$ =
200~GeV/62.4~GeV Au+Au at RHIC-PHENIX}
\def\leftmark{T.~Isobe for the PHENIX Collaboration}
 
\title{Measurement of neutral pions in $\sqrt{s_\mathrm{NN}}$ = 200~GeV
and 62.4~GeV Au+Au collisions at RHIC-PHENIX}
\authors{ 
{Tadaaki Isobe$^{1,a}$ (for the PHENIX Collaboration) %
\index{Isobe, T.} 
}\\[2.812mm]
{\normalsize
\hspace*{-8pt}$^1$ University of Tokyo,\\ 
Tokyo, Japan\\[0.2ex] 
}}
 
\abstract{Neutral pions~($\pi^0$) in $\sqrt{s_\mathrm{NN}}$ = 200~GeV
and 62.4~GeV Au+Au collisions were measured at RHIC-PHENIX Year-4 Run. 
In $\sqrt{s_\mathrm{NN}}$ = 200~GeV, $\pi^0$ specta are measured up to
p$_\mathrm{T}$ = 20~GeV/$c$. 
A strong suppression by a factor of $\sim$ 5 is observed and stays 
almost constant up to 20~GeV/$c$.}
 
\makeindex
\begin{document}
 
\maketitle

\section{Introduction}\label{intro}
One of the most intriguing observations at RHIC is that the yield of
$\pi^0$ at high transverse momentum~(p$_\mathrm{T}$) in central
$\sqrt{s_\mathrm{NN}}$ = 200~GeV Au+Au collisions compared to the yield
in p+p collisions scaled by the number of underlying nucleon-nucleon
collisions in Au+Au is suppressed~\cite{bib1,bib2}.   
The observed suppression is interpreted as a consequence of
jet-quenching effects, that is, hard-scattered partons produced in the
initial stage loose a large fraction of their kinetic energy while
traversing the hot and dense matter.

There are models that provide quantitative predictions of the
suppression.  
Each model involves various effects in a dense matter: Cronin and
nuclear shadowing as initial state effect, or energy loss through gluon
radiation as final state effect.
In order to reveal the parton energy loss mechanism in the dense
matter, it is important to measure $\pi^0$ systematically in
different energy systems and p$_{T}$ regions.

\section{$\pi^0$ measurement at PHENIX Year-4 Run}\label{run4}
At RHIC Year-4 run, PHENIX recorded the integrated luminosity of 0.24
nb$^{-1}$ in $\sqrt{s_\mathrm{NN}}$=200~GeV Au+Au collisions and
9.1$\mu$b$^{-1}$ in $\sqrt{s_\mathrm{NN}}$=62.4~GeV Au+Au collisions,
which allow us to extend the measurement of $\pi^0$ to much higher
p$_\mathrm{T}$, and to measure $\pi^0$ at an intermediate CMS energy.

$\pi^0$ is measured with the PHENIX electromagnetic
calorimeter~(EMCal)~\cite{bib:emc} via two-photon decay mode.  
EMCal is used to measure position and energy of photons. 

\subsection{Result on $\sqrt{s_\mathrm{NN}}$ = 200~GeV Au+Au collisions}\label{200gev}  
 
\begin{figure}[htb]
\vspace*{-.3cm}
\insertplot{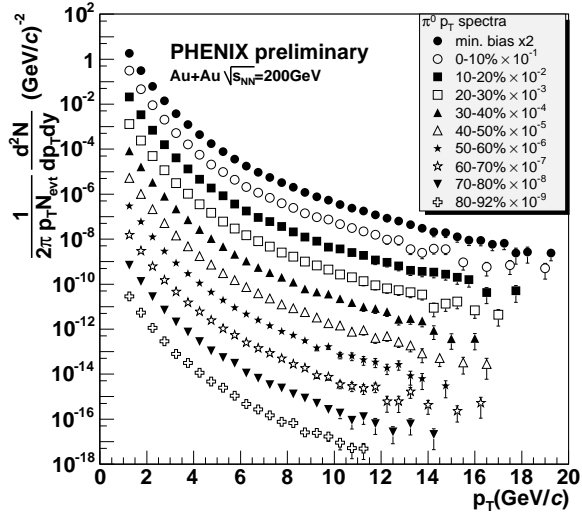}
\vspace*{-.8cm}
\caption[]{Invariant $\pi^0$ yields at y=0 versus p$_\mathrm{T}$ for minimum bias
 and 9 centralities in Au+Au at $\sqrt{s_\mathrm{NN}}$ = 200~GeV}
\label{fig1}
\vspace*{-.5cm}
\end{figure}

Figure~\ref{fig1} shows the preliminary data of fully corrected $\pi^0$  
invariant yield as a function of p$_\mathrm{T}$ for each centrality
in Au+Au at $\sqrt{s_\mathrm{NN}}$ = 200~GeV collisions.  
The large amount of data taken in RHIC Run4 has made it possible to
extend p$_\mathrm{T}$ region up to 20~GeV/$c$ for the central
collisions.
 
\begin{figure}[htb]
\vspace*{-.3cm}
\insertplot{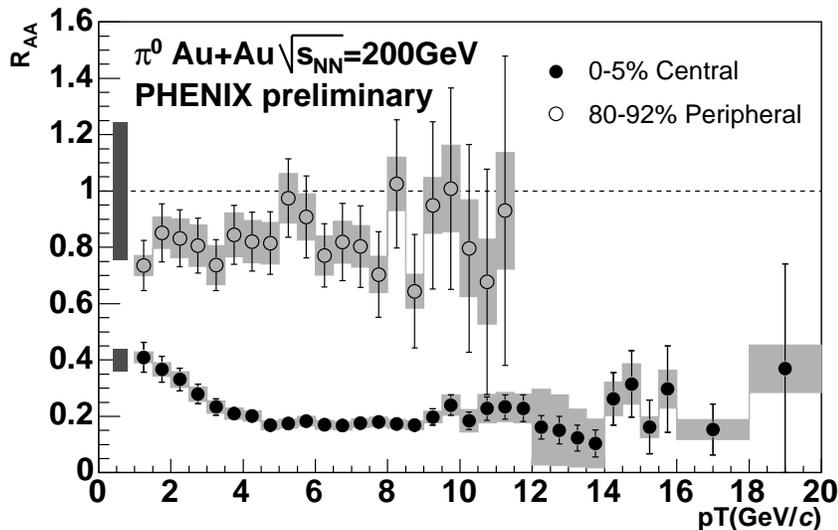}
\vspace*{-.7cm}
\caption[]{$\pi^0$ R$_\mathrm{AA}$ as a
 function of p$_\mathrm{T}$ for central (0-5\%) and peripheral (80-92\%)
 collisions. In addition to the statistical and pT-uncorrelated errors,
 point-to-point varying systematic errors are shown on the data points
 as boxes. And an overall systematic error of T$_\mathrm{AA}$
 normalization is shown at 1 and 0.4 for each centrality.}   
\label{fig2}
\vspace*{-.5cm}
\end{figure}

From fully corrected p$_\mathrm{T}$ spectra, the nuclear modification
factor~(R$_\mathrm{AA}$, see Appendix)
as a function of p$_\mathrm{T}$ is obtained.
PHENIX Run3 $\pi^0$ data is used as p+p reference for binary
scaling~\cite{bib:pppi0}.  
Figure~\ref{fig2} shows the R$_\mathrm{AA}$ for most central
events~(0-5\%) and peripheral events~(80-92\%). 
The suppression is very strong, and it is almost constant at 
R$_\mathrm{AA} \sim$ 0.2 up to very high p$_\mathrm{T}$.

\subsection{Result on $\sqrt{s_\mathrm{NN}}$ = 62.4~GeV Au+Au collisions}\label{62gev}

$\pi^0$ invariant yield at $\sqrt{s_\mathrm{NN}}$ = 62.4~GeV is measured
up to $\sim$ 7~GeV/$c$. 
For comparing with binary scaled p+p results, the CERN-ISR experimental 
results of $\pi^0$ cross section in p+p collisions are
used mainly, and details are described in \cite{bib:isrref}. 
There is discrepancy among each one of the CERN-ISR measurements, and
point-to-point systematic error is assigned as 25~\% for it.  

\begin{figure}[htb]
\vspace*{-.3cm}
\insertplot{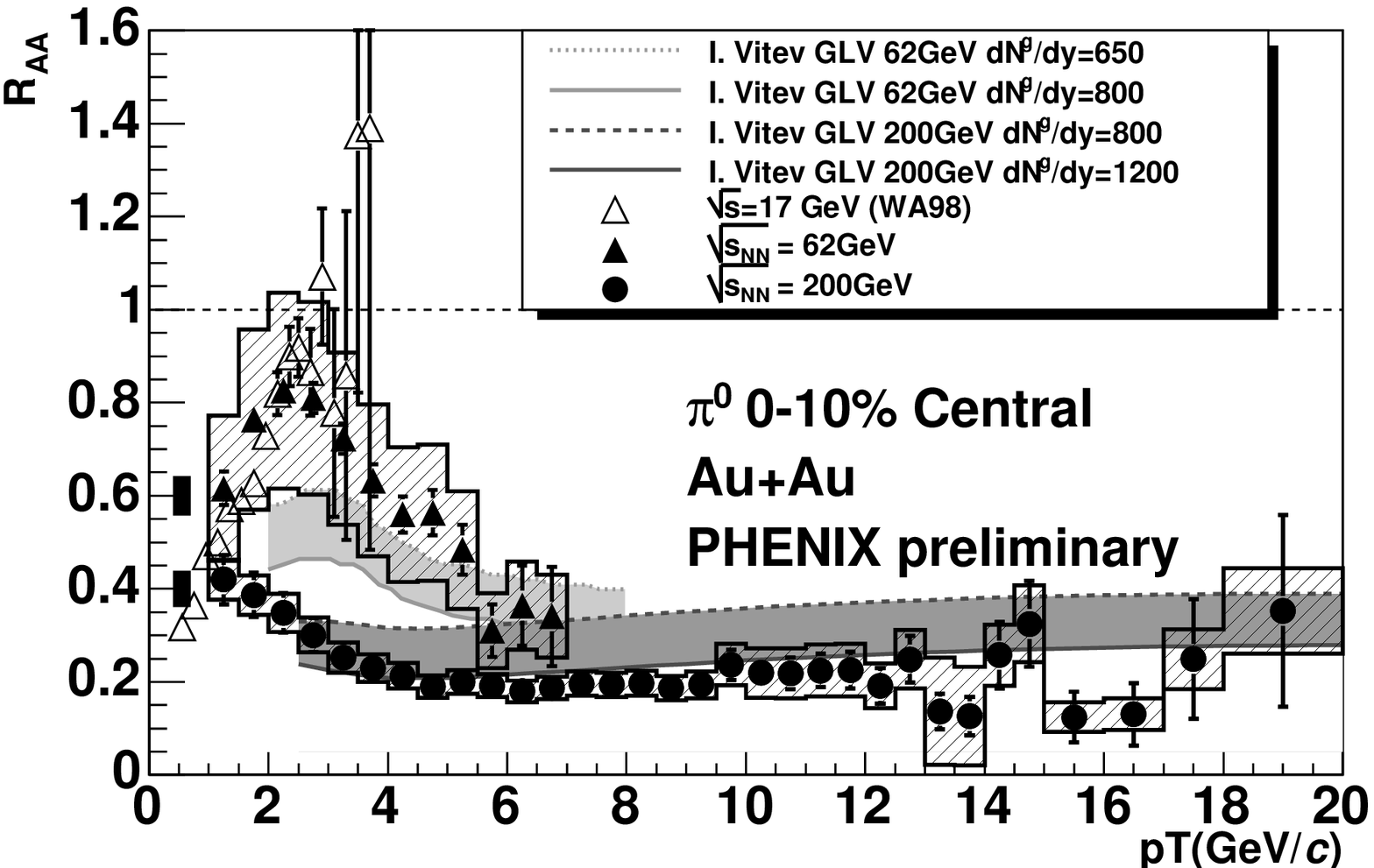}
\vspace*{-.7cm}
\caption[]{Comparison of $\pi^0$ R$_\mathrm{AA}$ PHENIX results with
 prediction of R$_\mathrm{AA}$ using the GLV model in
 $\sqrt{s_\mathrm{NN}}$ = 62.4~GeV and 200~GeV Au+Au collisions. 
 The 25\% systematic error due to the uncertainly of p+p
 reference is included in the point-to-point systematic error for the
 62.4~GeV data points.}  
\label{fig3}
\vspace*{-.5cm}
\end{figure}

Figure~\ref{fig3} shows the $\pi^0$ R$_\mathrm{AA}$ for
central events of $\sqrt{s_\mathrm{NN}}$ = 200~GeV and 62.4~GeV Au+Au
collisions compared with SPS $\sqrt{s_\mathrm{NN}}$ = 17~GeV Pb+Pb
R$_\mathrm{AA}$~\cite{bib:pbpb,bib:pbpbpp}, and theoretical prediction which
employs the GLV model~\cite{bib:62glv,bib:200glv}.  
The GLV model describes the strong suppression well and it indicates 
existence of bulk matter where dN$^g$/dy is more than 1100 in Au+Au
collisions at $\sqrt{s_\mathrm{NN}}$ = 200GeV.

\section{Conclusions}\label{concl}
$\pi^0$ is measured in $\sqrt{s_\mathrm{NN}}$ = 200~GeV and 62.4~GeV
Au+Au collisions.
The p$_\mathrm{T}$ region of $\pi^0$ spectra are extended up to
20~GeV/$c$ in $\sqrt{s_\mathrm{NN}}$ = 200~GeV.
A strong suppression by a factor of $\sim$ 5 is observed and stays 
almost constant up to 20~GeV/$c$.
 
\section*{Appendix}\label{app}
The degree of nuclear effect is quantified by the nuclear modification
factor~(R$_\mathrm{AA}$), defined by the following equation:

\begin{equation}
 R_\mathrm{AA}(\mathrm{p}_\mathrm{T}) =
  \frac{d^2N^\mathrm{AA}/d\mathrm{p}_\mathrm{T}d\eta}{T_\mathrm{AA}(b)d^2\sigma^\mathrm{NN}/d\mathrm{p}_\mathrm{T}d\eta}, \label{raa} 
\end{equation}

where the numerator is invariant $\pi^0$ yield in unit rapidity and
the denominator is the expected yield from p+p
collisions scaled by the number of binary nucleon-nucleon
collisions~($T_\mathrm{AA}(b) = N_\mathrm{coll}(b)/\sigma_\mathrm{NN}$)
in Au+Au.   
If a reaction is hard scattering with no nuclear effect, the nuclear
modification factor is unity. 

 
\section*{Notes} 
\begin{notes}
\vspace*{-.2cm}
\item[a]
Address: Graduate School of Science, University of Tokyo,\\
7-3-1 Hongo, Bunkyo-ku, Tokyo 113-0033, Japan\\
E-mail: isobe@cns.s.u-tokyo.ac.jp
\vspace*{-.3cm}
\end{notes}

\vfill\eject

\begin{thebibliography}{99}  
  
\bibitem{bib1}S.S~Adler {\it et al.} [PHENIX Collaboration], Phys.\ Rev.\ Lett.  {\bf 91} (2003) 072301.
\bibitem{bib2}S.S~Adler {\it et al.} [PHENIX Collaboration], Phys.\
	Rev.\ Lett.  {\bf 91} (2003) 072303.
 
\bibitem{bib:emc}
L.~Aphecetche {\it el al.}, Nucl.\ Instr.\ and\ Meth.\ {\bf A499} (2003) 521.
\bibitem{bib:pppi0}S.S~Adler {\it et al.} [PHENIX Collaboration], to be
	published. 

  \bibitem{bib:isrref}
D.~d'Enterria, nucl-ex/0411049.
  \bibitem{bib:pbpb}
M.~M.~Aggarwal {\it et al.} [WA98 Collaboration], Eur.\ Phys.\ J.\ C {\bf 23} (2002) 225.
  \bibitem{bib:pbpbpp}
D.~d'Enterria, Phys.\ Lett.\ B {\bf 596} (2004) 32.
\bibitem{bib:200glv}
I.~Vitev {\it el al.}, Phys.\ Rev.\ Lett.\ {\bf 89} (2002) 252301.
\bibitem{bib:62glv}
I.~Vitev, Phys.\ Lett.\ {\bf B606} (2005) 303.



\end{thebibliography}
\end{document}